\documentclass[10pt]{iopart}

\usepackage{iopams}  
\usepackage{cite} 
\usepackage{hyperref} 
\hypersetup{colorlinks=true, urlcolor=blue, linkcolor=blue, 
citecolor=blue} 
\usepackage{graphics} 
\usepackage{graphicx} 
\usepackage{mathrsfs}
\expandafter\let\csname equation*\endcsname\relax
\expandafter\let\csname endequation*\endcsname\relax
\usepackage{amsmath}
\usepackage{amssymb}

\begin{document}

\title[Phase transitions and thermal entanglement ]{Phase transitions and thermal entanglement of the distorted Ising-Heisenberg spin chain: topology of multiple-spin exchange interactions in spin ladders}

\author{Hamid Arian Zad}

\address{Young Researchers and Elite Club, Mashhad Branch, Islamic Azad University, Mashhad, Iran}
\eads{\mailto{\normalfont \color{blue} arianzad.hamid@mshdiau.ac.ir}}

\author{Nerses Ananikian}%
\address{Alikhanyan National Science Laboratory, Alikhanian Br. 2, 0036 Yerevan, Armenia}%
\eads{\mailto{\normalfont \color{blue} ananik@mail.yerphi.am}}
\vspace{10pt}

\begin{abstract}
We consider a symmetric spin-1/2 Ising-XXZ double sawtooth spin ladder obtained from distorting a spin chain, with the XXZ interaction between the interstitial Heisenberg dimers (which are connected to the spins based on the legs via an Ising-type interaction), the Ising coupling between nearest-neighbor spins of the legs  and rungs spins, respectively, and additional cyclic four-spin exchange (ring exchange) in square plaquette of each block. The presented analysis supplemented by results of the exact solution of the model with infinite periodic boundary implies a rich ground state phase diagram. As well as the quantum phase transitions, the characteristics of some of the thermodynamic parameters such as heat capacity, magnetization and magnetic susceptibility are investigated. We here prove that among the considered thermodynamic and thermal parameters, solely the heat capacity is sensitive versus the changes of the cyclic four-spin exchange interaction. By using the heat capacity function, we obtain a singularity relation between the cyclic four-spin exchange interaction and the exchange coupling between pair spins on each rung of the spin ladder. All thermal and thermodynamic quantities under consideration should investigate by regarding those points which satisfy the singularity relation. The thermal entanglement within the Heisenberg spin dimers is investigated by using the concurrence, which is calculated from a relevant reduced density operator in the thermodynamic limit.
\end{abstract}
%
%
%
\section{Introduction} \label{sec:level1}
Quantum spin chains as the most well-studied quantum models of statistical mechanics exhibit a wide variety of attractive phenomena at zero temperature such as the quantum phase transitions, which are driven by the change of the external parameters at absolute zero temperature \cite{Kitaev,Dillenschneider,Sachdev}, and also at low temperatures \cite{Werlang} that are controlled by the behavior of their ground states and low lying excited states. Entanglement \cite{Wootters,Arnesen} is one of the most interesting of such quantum phenomena, as it is a fundamental resource in performing most tasks in quantum computation and quantum information processing \cite{Osterloh1,Horodecki2007,Horodecki2009,Nag}. The relations between the ground state phase transitions and the quantum entanglement have already been revealed for the classes of spin-1/2 Heisenberg models \cite{Dillenschneider,Latorre,Dai,Amico}.

Interestingly, magnetic and thermodynamic properties of several insulating magnetic materials can be dramatically described by one-dimensional Heisenberg spin models. For example, a natural mineral, azurite $Cu_3(CO_3)_2(OH)_2$ can be considered as one of the first experimental realizations of the distorted diamond chain model \cite{Rule,Honecker,Ivanov1}. Recently, a considerable attention has been attracted to the Ising-Heisenberg diamond chains. The exactly solvable Ising-Heisenberg spin models provide a reasonable quantitative description of the thermal and magnetic behaviors associated with the materials with ordered structures in the real world \cite{Antonosyan,Paulinelli,Lisnyi}. Equivalent to the diamond chains \cite{Gu,Rojas2,Ananikian,Abgaryan,Strecka1}, magnetic spin ladders  \cite{Koga,Strecka} are a class of low-dimensional materials with structural and physical properties between those of one-dimensional chains and two-dimensional planes \cite{Bohnen}. Meanwhile, exact solvable one-dimensional lattices sawtooth chain with Ising-Heisenberg model and also Hubbard model have investigated from theoretical \cite{Derzhko,Blundell,Buttner} and experimental \cite{Bacq,Ueland} point of view.

First McCarron {\it et al.} introduced quasi-one dimentional dimer chain material $Sr_{14}Cu_{24}O_{41}$ with adjacent spin ladder layers structure, identical to the ladder layers in material $SrCu_2O_3$, and layers consist of $CuO_2$ chains \cite{McCarron}  (more explanations has been carried out in \cite{Vuletic}). 
In the structure of a spin ladder, the vertices have unpaired spins that interact along the legs via coupling constant parameter $\mathcal{J}_\parallel$ and along the rungs via $\mathcal{J}_\perp$, but are isolated from equivalent sites on adjacent ladders.

 Before, scientists suggested that some strongly correlated electron systems are expected to exhibit ring exchange, hence, some stimulating and deep studies was done on cuprates \cite{Roger}, solid $^3He$ \cite{Hetherington,Arakelyan,Toader} and spin ladders \cite{Toader,Okamoto,Muller} with considering four-spin ring exchange parameter.  According to the analysis of the low-lying excitation spectrum of the p-d model, it was shown that the Hamiltonian describing material $CuO_2$ should possess a finite value of ring exchange \cite{Muller}. Phase diagram of a Heisenberg spin-1/2 ladder with cyclic four-spin exchange was investigated in \cite{Laeuchli}. N. B. Ivanov { \it et al.} in  \cite{Ivanov} studied the phase diagram of a symmetric spin-1/2 Heisenberg diamond chain with additional cyclic four-spin exchange interaction by numerical exact diagonalization results.
 
 In the present paper, we study the one-dimensional $S = 1/2$ double sawtooth spin ladder obtained from distorting a simple diamond-like spin chain containing $N^{\prime}$ triangular spin-1/2 cells with multiple-spin exchange interactions. With regard to the significant in real experimental materials, we would like mention that using photonic lattices and optical lattices \cite{Parker1,Jo,Weimann,Silva} can be constructed a double sawtooth spin ladder with special model in the real world. For instance, authors in Ref. \cite{Parker1} used a lattice-shaking technique for hybridizing Bloch bands in optical lattices, indeed by shaking a lattice they could introduce a strong effective spin interaction also the organization of large ferromagnetic domains. Hence, our motivation to study this model is introducing new models as double sawtooth spin ladder which can form strong spin interaction of the real materials (lattices of cold  \cite{Parker1} and ultracold atoms  \cite{Jo}) created by means of an experimental action such as lattice-shaking. 
Starting from the nearest-neighbor interstitial Heisenberg dimer coupled legs, we use the cluster expansion technique to calculate the phase transitions and the Gibbs free energy in the parameter spaces, anisotropic Heisenberg XXZ interaction ${J}_H$(${J}_x$,${J}_z=\Delta$), $\mathcal{J}_\parallel$ and $\mathcal{J}_\perp$, and study the influence of the cyclic four-spin exchange on the thermodynamic parameters and spectrum.

The organization of this paper is as follows. In the next section, the solvable model and basic steps of the exact method are expressed. A brief discussion of the $T = 0$ phase diagram and the ground state properties of the model will be clarified in section \ref{QPT}. In section \ref{TMT}, the transfer-matrix solution of the system is given. Using the transfer-matrix formalism, the exact thermodynamic solution of the system is discussed. Plots of the heat capacity, the magnetization  and the magnetic susceptibility are presented in this section. In section  \ref{Concurrence}  we discuss the thermal entanglement and its magnetic phase transitions via concurrence. We end the paper with a brief conclusion in section \ref{conclusion}.
\section{Model}\label{Model}
\begin{figure}
\begin{center}
\resizebox{0.4\textwidth}{!}{%
 \includegraphics{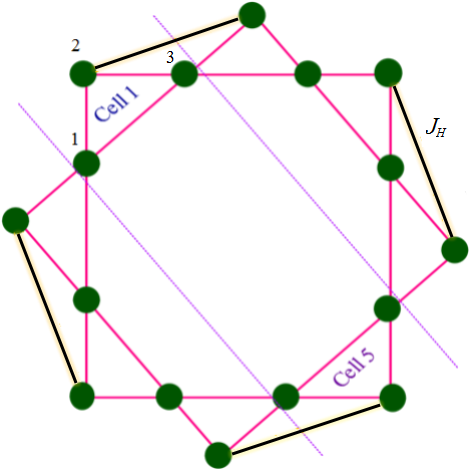}

  }
\caption{A diamond-like spin chain with Ising-XXZ Heisenberg model with Hamiltonian (\ref{DChamiltonian}) and $N=16$.}
\label{fig:figure1}  
\end{center}     
\end{figure}

\begin{figure}
\begin{center}
\resizebox{0.45\textwidth}{!}{%
  \includegraphics{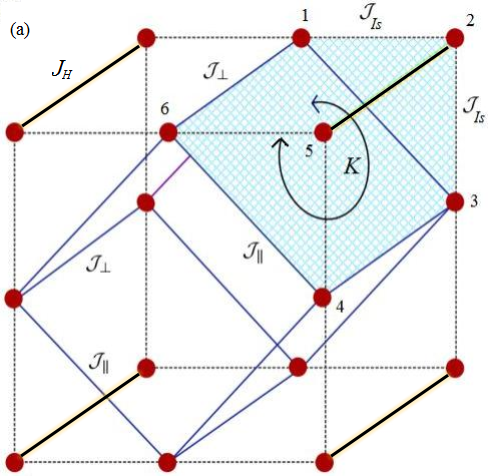}
}
\resizebox{0.25\textwidth}{!}{%
 \includegraphics{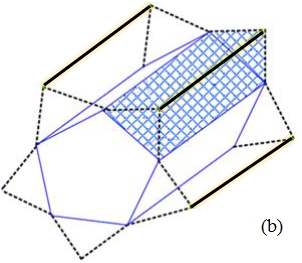}
  }
\caption{A $N$-spin double sawtooth ladder with Ising-XXZ Heisenberg model with Hamiltonian (\ref{SLhamiltonian}) obtained by distorting the diamond-like chain illustrated in figure \ref{fig:figure1}. (a) $N=16$, and (b) $N=20$.}
\label{fig:figure2}  
\end{center}     
\end{figure}

In this work, we study the Ising-XXZ Heisenberg double sawtooth ladder with mixed nodal Ising spins on the legs and interstitial dimer Heisenberg spins in the presence of an external magnetic field. The ladder under consideration is obtained from distorting a simple diamond-like spin chain contained triangular cells with periodic boundary conditions for which afront cells on the spin chain are connected together. The diamond-like spin chain and the corresponding spin ladder obtained from distortion are schematically illustrated in figures \ref{fig:figure1} and \ref{fig:figure2}, respectively. About the number of spins in the chain, it is noteworthy that due to the satisfying symmetry and defining an exact solvable model with long length, the number of spins are selected even and our method is used for $N\geq 8$.
 Consequently, to create a larger chain ($N>8$), the number of spins will grow as $N+4$, i.e., $N\in\{8,12,16,20,\cdots\}$. For instance, figure \ref{fig:figure1} shows a diamond-like spin chain with $N=16$, where the corresponding double sawtooth ladder is presented in figure \ref{fig:figure2}(a). To gain larger spin chain, the number of spins should be taken as $N=20$ (figure \ref{fig:figure2}(b) shows the corresponding double sawtooth ladder) and so on. The Hamiltonian operator of the diamond-like chain (figure \ref{fig:figure1}) can be expressed as
\begin{equation}\label{DChamiltonian}
\begin{array}{lcl}
H_{D}=\sum\limits_{i=1}^{N^{\prime}/2}{J}_H\textbf{S}_{2i-1,2}\cdot\textbf{S}_{2i,2}+ 
\sum\limits_{i=1}^{N^{\prime}}\mathcal{J}_{\parallel} \textbf{S}_{i,1}\cdot\textbf{S}_{i,3}+\mathcal{J}_{\perp}\big(\textbf{S}_{i,1}\cdot\textbf{S}_{i+\frac{N}{2},1}+\textbf{S}_{i,3}\cdot\textbf{S}_{i+\frac{N}{2},3}\big) \\
+\sum\limits_{j=1}^{N}\textbf{B}_{j}\cdot\textbf{S}_{j},
\end{array}
\end{equation}
where $N$ and $N^{\prime}$ are  the total number of spins and the number of triangular cells in the chain, respectively.

The form of the distorted double sawtooth ladder Hamiltonian containing two spins connected to the spins based on the legs so called interstitial Heisenberg dimer (figure \ref{fig:figure2}), under periodic boundary conditions such that the site $N+1$ would become equal to the first site, is given by the following SU(2) invariant $S=1/2$ model:
\begin{equation}\label{SLhamiltonian}
\begin{array}{lcl}
H_{SL}=\sum\limits_{i=1}^M{J}_H\textbf{S}^{\prime}_{i,2}\cdot\textbf{S}^{\prime}_{i,5}+
\sum\limits_{i=1}^M\mathcal{J}_{\parallel} \big({S}^{z\prime}_{i,1}\cdot{S}^{z\prime}_{i,3}+{S}^{z\prime}_{i,4}\cdot{S}^{z\prime}_{i,6}\big)+ \\
\sum\limits_{i=1}^M\mathcal{J}_{Is} \big({S}^{z\prime}_{i,1}\cdot{S}^{z\prime}_{i,2}+{S}^{z\prime}_{i,2}\cdot{S}^{z\prime}_{i,3}+{S}^{z\prime}_{i,4}\cdot{S}^{z\prime}_{i,5}+{S}^{z\prime}_{i,5}\cdot{S}^{z\prime}_{i,6}\big)+\\
\sum\limits_{i=1}^M\frac{\mathcal{J}_{\perp}}{2}\big({S}^{z\prime}_{i,1}\cdot{S}^{z\prime}_{i,6}+{S}^{z\prime}_{i,3}\cdot{S}^{z\prime}_{i,4}\big) +
K\sum\limits_{\langle 1346\rangle i}\big(P_{1346}^{i\circlearrowright}+P_{1346}^{i\circlearrowleft}\big)\\
-\sum\limits_{i=1}^{4M}\frac{{B}^{\prime}_z}{2}\big({S}^{z\prime}_{i,1}+{S}^{z\prime}_{i,3}+{S}^{z\prime}_{i,4}+{S}^{z\prime}_{i,6}\big)-\sum\limits_{i=1}^{4M}{B}^{\prime\prime}_z\big({S}^{z\prime}_{i,2}+{S}^{z\prime}_{i,5}\big),
\end{array}
\end{equation}
 where $M$ is the number of new transformed cells (shaded region in figure \ref{fig:figure2} as a block of the Hamiltonian (\ref{SLhamiltonian})), ${S}^{\prime}_{i,\gamma}$ ($\gamma=\{1,2,3,4,5,6\}$) are new labeled spins of transformed spin ladder, and ${J}_H$ is the exchange interaction between spins of the interstitial Heisenberg dimer on each block and 
 \begin{equation}\label{Ringoperator}
P_{1346}^{\circlearrowright}
\left(
\begin{array}{cc}
{S}^{z\prime}_{i,1} & {S}^{z\prime}_{i,3} \\
{S}^{z\prime}_{i,6} & {S}^{z\prime}_{i,4} \\
\end{array} \right) 
 =
 \left(
 \begin{array}{cc}
{S}^{z\prime}_{i,6} & {S}^{z\prime}_{i,1} \\
{S}^{z\prime}_{i,4} & {S}^{z\prime}_{i,3} \\
 \end{array}
 \right), 
 P_{1346}^{\circlearrowleft}
\left(
\begin{array}{cc}
{S}^{z\prime}_{i,1} & {S}^{z\prime}_{i,3} \\
{S}^{z\prime}_{i,6} & {S}^{z\prime}_{i,4} \\
\end{array} \right) 
 =
 \left(
 \begin{array}{cc}
 {S}^{z\prime}_{i,3} & {S}^{z\prime}_{i,4} \\
{S}^{z\prime}_{i,1} & {S}^{z\prime}_{i,6} \\
 \end{array}
 \right).
\end{equation}
  $\mathcal{J}_{\perp}$ and $\mathcal{J}_{\parallel}$ are the bilinear exchange couplings on the rungs and along the legs of the ladder, respectively, and $K$ is the coupling of the cyclic four-spin permutation operator per square plaquette existing in each block. $\mathcal{J}_{Is}$ is the Ising coupling between the spins on the legs of the square plaquette and two spins of the interstitial Heisenberg dimer.\\
  $2\textbf{S}=2\textbf{S}^{\prime}=\lbrace {\sigma}^x, {\sigma}^y, {\sigma}^z \rbrace$ are the spin operators (with $\hbar=1$).
 ${B}^{\prime}_z$ and ${B}^{\prime\prime}_z$ are applied homogeneous magnetic fields in the $z$-direction.
Note that here, all of introduced parameters are considered dimensionless.
First part of the Hamiltonian $H_{SL}$ is introduced as 
\begin{equation}\label{HamiltonianT}
\begin{array}{lcl}
\textbf{S}^{\prime}_{i,2}\cdot\textbf{S}^{\prime}_{i,5}={J}_{x} \big({S}^{x\prime}_{i,2}{S}^{x\prime}_{i,5}+{S}^{y\prime}_{i,2}{S}^{y\prime}_{i,5}\big)+\Delta{S}^{z\prime}_{i,2}{S}^{z\prime}_{i,5}.\\
\end{array}
\end{equation}
We rewrite the operator $P_{1346}^i$ as a product of two spin permutation operators and obtain the following result which contains both bilinear and biquadratic terms of the spin-1/2 operators
\begin{equation}\label{HamiltonianRing}
\begin{array}{lcl}
\mathcal{H}_{ring}=\\
\frac{K}{2}\sum\limits_{\langle 1346\rangle i}\big[{S}^{z\prime}_{i,1}\cdot{S}^{z\prime}_{i,3}+{S}^{z\prime}_{i,3}\cdot{S}^{z\prime}_{i,4}+
{S}^{z\prime}_{i,4}\cdot{S}^{z\prime}_{i,6}+{S}^{z\prime}_{i,6}\cdot{S}^{z\prime}_{i,1}+\frac{1}{4}\big]\\
+2K\sum\limits_{\langle 1346\rangle i}({S}^{z\prime}_{i,1}\cdot{S}^{z\prime}_{i,3})({S}^{z\prime}_{i,4}\cdot{S}^{z\prime}_{i,6})+({S}^{z\prime}_{i,1}\cdot{S}^{z\prime}_{i,6})({S}^{z\prime}_{i,3}\cdot{S}^{z\prime}_{i,4})\\
-({S}^{z\prime}_{i,1}\cdot{S}^{z\prime}_{i,4})({S}^{z\prime}_{i,3}\cdot{S}^{z\prime}_{i,6}).
\end{array}
\end{equation}

The Hamiltonian ${h}_{i}$ involves all the interaction terms of the i-th unit block and it can be written as
\begin{equation}\label{HamiltonianT}
\begin{array}{lcl}
{h}_{i}=\big[{J}_{x}\big({S}^{x\prime}_{i,2}{S}^{x\prime}_{i,5}+{S}^{y\prime}_{i,2}{S}^{y\prime}_{i,5}\big)+\Delta{S}^{z\prime}_{i,2}{S}^{z\prime}_{i,5}\big]\\
+\mathcal{J}_{\parallel}\big({S}^{z\prime}_{i,1}{S}^{z\prime}_{i,3}+{S}^{z\prime}_{i,4}{S}^{z\prime}_{i,6}\big)+\frac{\mathcal{J}_{\perp}}{2}\big({S}^{z\prime}_{i,1}{S}^{z\prime}_{i,6}+{S}^{z\prime}_{i,3}{S}^{z\prime}_{i,4}\big)\\
+\mathcal{J}_{Is}\big[{S}^{z\prime}_{i,1}{S}^{z\prime}_{i,2}+{S}^{z\prime}_{i,2}{S}^{z\prime}_{i,3}+{S}^{\prime}_{i,4}{S}^{z\prime}_{i,5}+
{S}^{\prime}_{i,5}{S}^{z\prime}_{i,6}\big]\\
-\frac{{B}^{\prime}_{z}}{2}\big({S}^{z\prime}_{i,1}+{S}^{z\prime}_{i,3}+{S}^{z\prime}_{i,4}+{S}^{z\prime}_{i,6}\big)-{B}^{\prime\prime}_{z}\big({S}^{z\prime}_{i,2}+{S}^{z\prime}_{i,5}\big)+\\
\mathcal{H}_{ring}.
\end{array}
\end{equation}
In this case, $\mathcal{J}, J_H>0$ is denoted ferromagnetic exchange interactions and $\mathcal{J},J_H<0$ antiferromagnetic ones.
 $\mathcal{H}_{ring}$ denotes the Hamiltonian of the ring (\ref{HamiltonianRing}) with Ising interaction. The longitudinal external magnetic field $B^{\prime\prime}_z$ acts on the Heisenberg dimer spins and the magnetic field $B^{\prime}_z$ acts on Ising spins of the Hamiltonian. For simplicity, we will consider the case $B^{\prime}_z=2B^{\prime\prime}_z=2B$.
\section{The zero temperature phase diagram}\label{QPT}
Let us explain possible zero temperature ground states of the spin-1/2 Ising-Heisenberg double sawtooth ladder with Ising four-spin ring exchanges, which can be written as a tensor product over the lowest-energy eigenstates of the block Hamiltonian (\ref{HamiltonianT}) under the corresponding energies per block. One can find by inspection different ground states, namely, ordered ferrimagnetic phase (CFI), classical ferromagnetic (CFM), entangled antiferromagnetic phase (EAFM), and two different entangled ferrimagnetic phases (EFI 1 and EFI 2). Spin arrangements of the relevant phases can be  demonstrated by using the following tensor product of the eigenvectors
\begin{equation}\label{PhaseD}
\begin{array}{lcl}
\vert CFM\rangle=\displaystyle\prod_{i=1}^{M}\vert \uparrow\uparrow\uparrow\uparrow\rangle_i\otimes\vert\varphi_1\rangle_i,\\
\vert EAFM\rangle=\displaystyle\prod_{i=1}^{M}\vert\uparrow\uparrow\downarrow\downarrow \rangle_i\otimes\vert\varphi_2\rangle_i,\\
\vert EFI 1\rangle=\displaystyle\prod_{i=1}^{M}\vert \uparrow\uparrow\uparrow\downarrow\rangle_i\otimes\vert\varphi_2\rangle_i,\\
\vert CFI\rangle=\displaystyle\prod_{i=1}^{M}\vert \downarrow\downarrow\downarrow\downarrow\rangle_i\otimes\vert\varphi_1\rangle_i,\\
\vert EFI 2\rangle=\displaystyle\prod_{i=1}^{M}\vert \uparrow\uparrow\uparrow\uparrow\rangle_i\otimes\vert\varphi_3\rangle_i,\\
\end{array}
\end{equation}
where
\begin{equation}\label{PhaseD}
\begin{array}{lcl}
\vert \varphi_1\rangle_i=\vert\uparrow\uparrow\rangle_i,\\
\vert \varphi_2\rangle_i=\frac{1}{2}(\vert\uparrow\downarrow\rangle+\vert\downarrow\uparrow\rangle)_i,\\
\vert \varphi_3\rangle_i=\frac{1}{2}(\vert\uparrow\downarrow\rangle-\vert\downarrow\uparrow\rangle)_i,\\
\vert \varphi_4\rangle_i=\vert\downarrow\downarrow\rangle_i,\\
\end{array}
\end{equation}
are the eigenstates of the interstitial XXZ Heisenberg dimer, and $\vert 1346\rangle_i$ is the state of the square plaquette in $i$-th block.
We consider these ground states because of their amazing behavior under the changes of the used parameters specially the cyclic four-spin exchange $K$. 

\begin{figure}
\begin{center}
\resizebox{0.5\textwidth}{!}{%
  \includegraphics{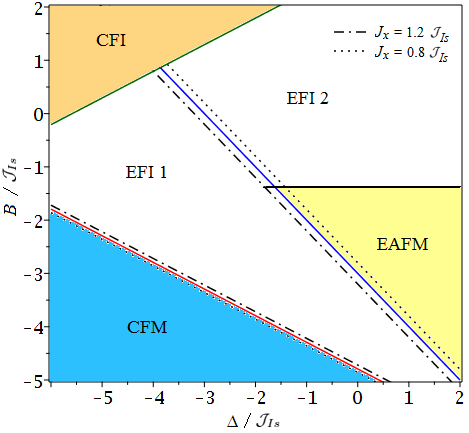}
 }
\caption{The ground-state phase diagram of the spin-1/2 Ising-Heisenberg double sawtooth ladder in the $B-\Delta$ plane at fixed values of ${J}_{x}=\mathcal{J}_{\parallel}=K=\mathcal{J}_{Is}$ and $\mathcal{J}_{\perp}=4\mathcal{J}_{Is}$.}
\label{fig:QPTBDelta}       
\end{center}
\end{figure}

Figure \ref{fig:QPTBDelta} shows phase diagram at zero temperature as a dependency of $B$ and $\Delta$ for the fixed values of $\mathcal{J}_{\parallel}=\mathcal{J}_{Is}$,  $\mathcal{J}_{\perp}=4\mathcal{J}_{Is}$ and $K=\mathcal{J}_{Is}$. For $B\lesssim -1.4$ there are two boundaries between CFM and EFI 1, also between EFI 1 and EAFM which are represented by the lines $B=-\Delta-3$ and $B=-1/2(\Delta+11-\sqrt{2})$, respectively. For $B\gtrsim -1.4$ there are three regions related to the phases EFI 1, CFI and EFI 2 which are separated with lines $B=1/2(\Delta+7-\sqrt{2})$ and $B=-1/2(\Delta+11-\sqrt{2})$. It is noteworthy that in this case we consider a fixed value for $K$, though we investigated the phase diagram in the $B-\Delta$ plane with respect to the various values of $K$ and concluded that the changes of this parameter is effectless. On the other hand, we changed the value of ${J}_{x}$ for boundaries between CFM and EFI 1, and between EFI 1 and EAFM and concluded that this parameter has essential effect on the mentioned boundaries, namely, with increase (decrease) of ${J}_{x}$ the region of EFI 1 phase is restricted (spreading). Dashed-dot and dot lines depict these changes.

\begin{figure}
\begin{center}
\resizebox{0.6\textwidth}{!}{%
\includegraphics{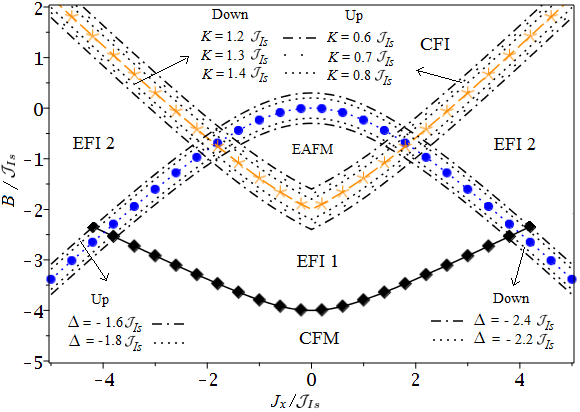}
}
\caption{The ground-state phase diagram of the spin-1/2 Ising-Heisenberg double sawtooth ladder in the $B-{J}_{x}$ plane at fixed values of $\mathcal{J}_{\parallel}=\mathcal{J}_{Is}$,  $\mathcal{J}_{\perp}=4\mathcal{J}_{Is}$.}
\label{fig:QPTBJ}
\end{center}
\end{figure}

In figure \ref{fig:QPTBJ} we plot the phase diagram $B/\mathcal{J}_{Is}$ versus ${J}_{x}/\mathcal{J}_{Is}$ at zero temperature and negative values of $\Delta$ for $\mathcal{J}_{\parallel}=\mathcal{J}_{Is}$ and $\mathcal{J}_{\perp}=4\mathcal{J}_{Is}$, where the ground-state energies for $\Delta=-2\mathcal{J}_{Is}$ are given by
\begin{equation}\label{PF}
\begin{array}{lcl}
\mathcal{E}_{CFM}=\frac{65}{4}+6\frac{B}{\mathcal{J}_{Is}}, \\
\mathcal{E}_{EAFM}=\frac{17}{4}+2\sqrt{(\frac{{J}_{x}}{\mathcal{J}_{Is}})^2+4}, \\
\mathcal{E}_{CFI}=\frac{33}{4}-2\frac{B}{\mathcal{J}_{Is}}, \\
\mathcal{E}_{EFI 1}=-\frac{7}{4}+2\frac{B}{\mathcal{J}_{Is}}+2\sqrt{(\frac{{J}_{x}}{\mathcal{J}_{Is}})^2+1}, \\
\mathcal{E}_{EFI 2}=-\frac{65}{4}+4\frac{B}{\mathcal{J}_{Is}}+2\vert\frac{{J}_{x}}{\mathcal{J}_{Is}}\vert, \\
 \end{array}
\end{equation}
This figure shows that the phase boundary between the CFM and EFI 1 states is limited by the curve $\frac{B}{\mathcal{J}_{Is}}=-\frac{9}{2}+\frac{1}{2}\sqrt{(\frac{{J}_{x}}{\mathcal{J}_{Is}})^2+1}$, whereas between the EFI 1 and EAFM states it is limited by the curve  $\frac{B}{\mathcal{J}_{Is}}=-3+\frac{1}{2}\vert\frac{{J}_{x}}{\mathcal{J}_{Is}}\vert+\frac{1}{2}\sqrt{(\frac{{J}_{x}}{\mathcal{J}_{Is}})^2+4}$. Regions of the phases EAFM and CFI are separated with the curve $\frac{B}{\mathcal{J}_{Is}}=2-\frac{1}{2}\sqrt{(\frac{{J}_{x}}{\mathcal{J}_{Is}})^2+4}$. The state EFI 2 is limited by EFI 1 region, where the same curve $\frac{B}{\mathcal{J}_{Is}}=2-\frac{1}{2}\sqrt{(\frac{{J}_{x}}{\mathcal{J}_{Is}})^2+4}$ can also depict the phase boundary between them.

We investigated the phase  boundary between the EFI 1 and EAFM states in the $B-{J}_{x}$ plane with respect to the various values of $K$ and concluded that this parameter has essential role to determine the region of the both states. Indeed by increasing the cyclic four-spin exchange $K$( from fixed value $K=1$), the region of the EAFM state increases, then it makes the EFI 1 region be limited. When $K$ decreases the EAFM region decreases, and in turn the EFI 1 region becomes wider. According to our investigations, under the considered conditions the changes of $\Delta$ have no effect on the  boundary between the  EFI 1 and EAFM states, while the changes of $K$ can move the boundary phase between these states and also the EFI 2 and CFI. On the other hand, the changes of $K$ have no effect on the boundary between the CFI and EAFM states, and also between the EFI 1 and EFI 2 states (blue circles), while $\Delta$ can extremely effect on these boundaries. We show explicitly these effects with different lines in the figure.

\section{Solution within the transfer-matrix technique}\label{TMT}
We use the transfer matrix technique for describing the thermodynamics of the system with new transformed block Hamiltonian (\ref{HamiltonianT}) under consideration. The interstitial Heisenberg dimer coupling can be expressed as
\begin{equation}\label{density matrices}
\big(\textbf{S}^{\prime}_{i,2}\cdot\textbf{S}^{\prime}_{i,5}\big)_{\Delta,{J}_{x}}= \left(
\begin{array}{cccc}
\frac{\Delta}{4} & 0 & 0 & 0 \\
0 & -\frac{\Delta}{4} & \frac{{J}_x}{2} & 0 \\
0 &\frac{{J}_x}{2} & -\frac{\Delta}{4}  & 0\\
 0 & 0 & 0 & \frac{\Delta}{4}
\end{array} \right).
\end{equation}
Due to the commutation relation between different block Hamiltonians, $[{h}_{i},{h}_{j}]=0$, the partition function of the Ising-Heisenberg spin ladder can be written in the form
\begin{equation}\label{PF}
\mathcal{Z}=Tr\big[\exp(-\beta H_{SL})\big]=Tr\Big[\displaystyle\prod_{i=1}^{M}\exp(-\beta h_{i})\Big],
\end{equation}
where $\beta=\frac{1}{k_{B}T}$, $k_{B}$ is the Boltzmann’s constant and $T$ is the absolute temperature.
We can consider the following matrix representation for $\exp(-\beta h_{i})$ in the two qubit standard basis of the eigenstates of the composite spin operators $\{ S_{i,1}^{z\prime},S_{i,6}^{z\prime},S_{i,3}^{z\prime},S_{i,4}^{z\prime} \} $ of the two consecutive rungs of the square plaquette in block $i$, by which the partition function $\mathcal{Z}$ can be defined as
\begin{equation}\label{PF}
\begin{array}{lcl}
\mathcal{Z}=Tr\big[ \langle S_{1,1}^{z\prime}S_{1,6}^{z\prime}\vert\mathcal{T}\vert S_{1,3}^{z\prime}S_{1,4}^{z\prime}\rangle\langle S_{2,1}^{z\prime}S_{2,6}^{z\prime}\vert \mathcal{T}\vert S_{2,3}^{z\prime}
S_{2,4}^{z\prime}\rangle\cdots\langle S_{M,1}^{z\prime}S_{M,6}^{z\prime}\vert \mathcal{T}\vert S_{M,3}^{z\prime}S_{M,4}^{z\prime}\rangle \big],
\end{array}
\end{equation}
where $S_{i,j}^{z\prime}=\pm\frac{1}{2}$ and
\begin{equation}\label{TrM}
\begin{array}{lcl}
\mathcal{T}(i)=\langle S_{i,1}^{z\prime}S_{i,6}^{z\prime}\vert\exp(-\beta h_{i})\vert S_{i,3}^{z\prime}S_{i,4}^{z\prime}\rangle=
\sum\limits_{k=1}^4\exp\big[-\beta\mathcal{E}_k(S_{i,1}^{z\prime}S_{i,6}^{z\prime},S_{i,3}^{z\prime}S_{i,4}^{z\prime})\big].
\end{array}
\end{equation}
Four eigenvalues of the $i-$th block with Hamiltonian $h_{i}$ are
\begin{equation}\label{eigenvalues1}
\begin{array}{lcl}
\mathcal{E}_1(i)=\frac{\Delta}{4}+\mathcal{J}_{Is}\big(S_{i,1}^{z\prime}+S_{i,3}^{z\prime}+S_{i,4}^{z\prime}+S_{i,6}^{z\prime}\big)+\Xi+B,\\
\mathcal{E}_2(i)=\\ 
-\frac{\Delta}{4}+\Xi+\frac{1}{2}\sqrt{4\mathcal{J}_{Is}^2\big(S_{i,1}^{z\prime}+S_{i,3}^{z\prime}-S_{i,4}^{z\prime}-S_{i,6}^{z\prime}\big)^2+{J}_{x}^2},\\
\mathcal{E}_3(i)=\\ 
-\frac{\Delta}{4}+\Xi-\frac{1}{2}\sqrt{4\mathcal{J}_{Is}^2\big(S_{i,1}^{z\prime}+S_{i,3}^{z\prime}-S_{i,4}^{z\prime}-S_{i,6}^{z\prime}\big)^2+{J}_{x}^2},\\
\mathcal{E}_4(i)=\frac{\Delta}{4}-\mathcal{J}_{Is}\big(S_{i,1}^{z\prime}+S_{i,3}^{z\prime}+S_{i,4}^{z\prime}+S_{i,6}^{z\prime}\big)+\Xi-B,\\
\end{array}
\end{equation}
where 
\begin{equation}\label{Xi}
\begin{array}{lcl}
\Xi=\mathcal{J}_{\parallel}\big(S_{i,1}^{z\prime}S_{i,3}^{z\prime}+S_{i,4}^{z\prime}S_{i,6}^{z\prime}\big)+\frac{\mathcal{J}_{\perp}}{2}\big(S_{i,1}^{z\prime}S_{i,6}^{z\prime}+S_{i,3}^{z\prime}S_{i,4}^{z\prime}\big)+\\
K\big(S_{i,1}^{z\prime}S_{i,3}^{z\prime} +S_{i,3}^{z\prime}S_{i,4}^{z\prime}+S_{i,4}^{z\prime}S_{i,6}^{z\prime}+S_{i,6}^{z\prime}S_{i,1}^{\prime}+
4S_{i,1}^{z\prime}S_{i,3}^{z\prime}S_{i,4}^{z\prime}S_{i,6}^{z\prime}+\frac{1}{4}\big)+\\
B\big(S_{i,1}^{z\prime}+S_{i,3}^{z\prime}+S_{i,4}^{z\prime}+S_{i,6}^{z\prime}\big).
\end{array}
\end{equation}
We can figure out the transfer matrix $\mathcal{T}(i)$ as follows:
\begin{equation}\label{Tmat}
\mathcal{T}(i) = \left(
\begin{array}{cccc}
 \mathcal{A} & \alpha & \alpha & \Omega \\
\alpha & \mathcal{P} & \xi & \delta \\
 \alpha & \xi & \mathcal{P} & \delta\\
 \Omega & \delta & \delta & \mathcal{F}
\end{array} \right),
\end{equation}
where the elements of the transfer matrix are defined through eigenvalues (\ref{eigenvalues1}) as
\begin{equation}\label{TCoefficient}
\begin{array}{lcl}
 \mathcal{A} =2\exp\big[-\frac{1}{4}\beta(\Delta+8\mathcal{J}_{\parallel}+4\mathcal{J}_{\perp}+33K+16B)\big]\times\\
 \big(\cosh(\beta(B+2\mathcal{J}_{Is}))+\exp\big[\frac{1}{2}\beta\Delta\big]\cosh(\frac{1}{2}\beta{J}_x )\big)\\
 \alpha=2\exp\big[-\frac{1}{4}\beta(\Delta-15K+8B)\big]\times\\
 \big(\cosh(\beta(B+\mathcal{J}_{Is})) +\exp\big[\frac{1}{2}\beta\Delta\big]\cosh\big(\frac{1}{2}\beta\sqrt{{J}_x^2+4\mathcal{J}_{Is} ^2}\big)\big)
 \\
 \Omega=2\exp\big[-\frac{1}{4}\beta\big(\Delta+8\mathcal{J}_{\parallel}-4\mathcal{J}_{\perp}+17K\big)\big]\times\\
 \big(\cosh(\beta B)+ \exp\big[\frac{1}{2}\beta\Delta)\cosh\big(\frac{1}{2}\beta\sqrt{{J}_x^2+16\mathcal{J}_{Is} ^2}\big)\big)\\
 \mathcal{P}=2\exp\big[-\frac{1}{4}\beta(\Delta-8\mathcal{J}_{\parallel}-4\mathcal{J}_{\perp}+K)\big]\times\\
 \big(\cosh(\beta B)+\exp\big[\frac{1}{2}\beta\Delta\big]\cosh(\frac{1}{2}\beta{J}_x )\big)\\
\xi=2\exp\big[-\frac{1}{4}\beta(\Delta-8\mathcal{J}_{\parallel}+4\mathcal{J}_{\perp}+17K)\big]\times\\
\big(\cosh(\beta B)+\exp\big[\frac{1}{2}\beta\Delta\big]\cosh(\frac{1}{2}\beta{J}_x)\big)\\
\delta=2\exp\big[-\frac{1}{4}\beta(\Delta+15K+8B)\big]\times\\
 \big(\cosh(\beta(B-\mathcal{J}_{Is})) +\exp\big[\frac{1}{2}\beta\Delta\big]\cosh\big(\frac{1}{2}\beta\sqrt{{J}_x^2+4\mathcal{J}_{Is} ^2}\big)\big)\\
\mathcal{F}=2\exp\big[-\frac{1}{4}\beta(\Delta+8\mathcal{J}_{\parallel}+4\mathcal{J}_{\perp}+33K-16B)\big]\times\\
 \big(\cosh(\beta(B-2\mathcal{J}_{Is}))+\exp\big[\frac{1}{2}\beta\Delta\big]\cosh(\frac{1}{2}\beta{J}_x )\big).
\end{array}
\end{equation}
As the spin ladder is translational invariant and all of $h_i$ are  independent of the site $i$, equation (\ref{PF}) can be expressed as
\begin{equation}\label{ConvertedZ}
\mathcal{Z}=Tr\big[\mathcal{T}^M\big],
\end{equation}
hence the total partition function may be expressed in terms of four eigenvalues of the transfer matrix $\mathcal{T}$:
\begin{equation}\label{TotalZ}
\mathcal{Z}=\Lambda_1^{M}+\Lambda_2^{M}+\Lambda_3^{M}+\Lambda_4^{M}.
\end{equation}
 In the thermodynamic limit, it is sufficient to consider only the largest eigenvalue $\Lambda_{max}$ to calculate the partition function, hence, the free energy per block for infinitely long chain, when only the maximal eigenvalue survives is given by \cite{Paulinelli}
 \begin{equation}\label{FreeE}
 \begin{array}{lcl}
 f=f_0+f^{\prime}=
 2{J}_x+\Delta-\frac{1}{\beta}\lim\limits_{M\rightarrow \infty}\ln\frac{1}{M}\mathcal{Z}=2{J}_x+\Delta-\frac{1}{\beta}\ln\Lambda_{max},
 \end{array}
 \end{equation}
 where 
\begin{equation}\label{LamdaMax}
\begin{array}{lcl}
\Lambda_{max}=\\
4\sinh\big(\beta(\mathcal{J}_{\perp}+2K)\big)\exp\big[-\frac{1}{4}\beta(-8\mathcal{J}_{\parallel}+9K)\big]\times\\
\big(\exp\big[-\frac{1}{4}\beta\Delta\big]\cosh(\beta B) +\exp\big[\frac{1}{4}\beta\Delta\big]\cosh\big(\frac{1}{2}\beta {J}_x\big)\big).
\end{array}
\end{equation}
\subsection{Thermodaynamic parameters}\label{TM}
\begin{figure}
 \begin{center}
\resizebox{0.9\textwidth}{!}{%
\includegraphics{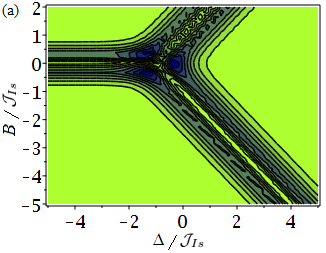} 
\includegraphics{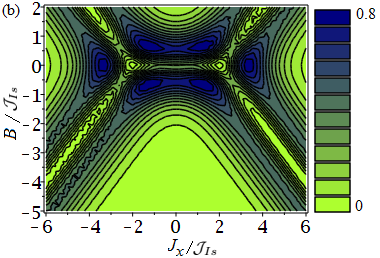}
}
\caption{Contour plots of the heat capacity as function of  the magnetic field versus (a) $\Delta/\mathcal{J}_{Is}$ at fixed values of
 ${J}_{x}=\mathcal{J}_{\parallel}=K=\mathcal{J}_{Is}$,  $\mathcal{J}_{\perp}=4\mathcal{J}_{Is}$ and $T=0.5\mathcal{J}_{Is}$, and (b) ${J}_{x}/ \mathcal{J}_{Is}$ at fixed values of $\mathcal{J}_{\parallel}=K=\mathcal{J}_{Is}$, 
  $\mathcal{J}_{\perp}=4\mathcal{J}_{Is}$, $\Delta=-2\mathcal{J}_{Is}$ and $T=0.5\mathcal{J}_{Is}$.}
\label{fig:heatcapacitycontour}
\end{center}
\end{figure}

The heat capacity, the magnetization and the magnetic susceptibility can be calculated by using the maximal eigenvalue of the transfer matrix through the following formulae
\begin{equation}
\begin{array}{lcl}
 \mathcal{C}=k_B\beta^{2}\frac{\partial^{2}( \ln\Lambda_{max})}{\partial^{2}\beta}, \mathcal{M}=-\frac{\partial f}{\partial B}, \chi=\frac{\partial\mathcal{M}}{\partial B}.
\end{array}
\end{equation}
General features of the heat capacity behavior are presented as contour plots in figure \ref{fig:heatcapacitycontour}. 
Figure \ref{fig:heatcapacitycontour}(a) shows field dependence of the heat capacity versus $\Delta/\mathcal{J}_{Is}$ at low temperature for the fixed values of exchange couplings  $\mathcal{J}_{\parallel}=\mathcal{J}_{Is}$,  $\mathcal{J}_{\perp}=4\mathcal{J}_{Is}$, ${J}_{x}=\mathcal{J}_{Is}$ and $K=\mathcal{J}_{Is}$. For the critical points starting from the EFI 1 phase to CFI and EFI 2 the heat capacity  becomes maximum. Indeed, this thermodynamic parameter divides the phase diagram of the system to three regions related to the phases, EFI 1, CFI and EFI 2 at the considered conditions. 
Figure \ref{fig:heatcapacitycontour}(b) represents contour plots of the heat capacity in the  $B-{J}_{x}$ plane at low temperature and fixed ${J}_{x}=\mathcal{J}_{\parallel}=\mathcal{J}_{Is}$,  $\mathcal{J}_{\perp}=4\mathcal{J}_{Is}$, $\Delta=-2\mathcal{J}_{Is}$ and $K=\mathcal{J}_{Is}$. By inspecting this figure, one can clearly find all regions related to the ground state phases CFM- EFI 1, CFI and EFI 2 (green regions) which the heat capacity is minimum. On the one hand,  when system is in EAFM state (blue region) the heat capacity is maximum. Consequently, the heat capacity of a spin ladder with infinite size obtained from largest eigenvalue of the transfer matrix, can be a very good candidate to determine phase boundaries between the ground states of the spin ladder under consideration.
\begin{figure}
\begin{center}
\resizebox{0.6\textwidth}{!}{%
\includegraphics{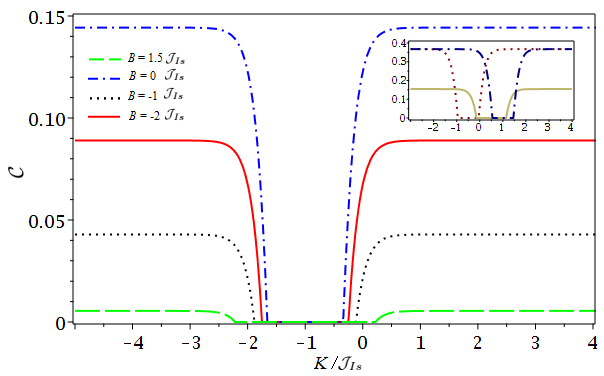}
}
\caption{The heat capacity ($\max(0,\mathcal{C})$) as a function of the cyclic four-spin exchange $K/\mathcal{J}_{Is}$ at low temperature ($T=0.5\mathcal{J}_{Is}$) and fixed $\mathcal{J}_{\parallel}={J}_{x}=\mathcal{J}_{Is}$, $\mathcal{J}_{\perp}=4\mathcal{J}_{Is}$, $\Delta=-2\mathcal{J}_{Is}$ for several values of the magnetic field $B$. Inset shows the heat capacity versus $K/\mathcal{J}_{Is}$ for the various values of $\mathcal{J}_{\perp}$ such that doted line corresponds to the case when $\mathcal{J}_{\perp}=\mathcal{J}_{Is}$, solid line when $\mathcal{J}_{\perp}=-\mathcal{J}_{Is}$, and dashed-dot line when $\mathcal{J}_{\perp}=-2\mathcal{J}_{Is}$, at fixed  $\mathcal{J}_{\parallel}=\mathcal{J}_{Is}$, $\Delta=-2\mathcal{J}_{Is}$, $T=0.5\mathcal{J}_{Is}$ and $B=0.5\mathcal{J}_{Is}$.}
\label{fig:2DHeatCapacityK}
\end{center}
\end{figure}

To gain an insight into the effect of cyclic four-spin exchange, thermal variations of the heat capacity $\mathcal{C}$ versus ratio 
$K/\mathcal{J}_{Is}$ at low temperature are plotted in figure \ref{fig:2DHeatCapacityK} for several values of the magnetic field $B/\mathcal{J}_{Is}$ with conditions ${J}_{x}=\mathcal{J}_{\parallel}=\mathcal{J}_{Is}$, $\mathcal{J}_{\perp}=4\mathcal{J}_{Is}$, and $\Delta=-2\mathcal{J}_{Is}$. 
In this case, we investigate $\max(0,\mathcal{C})$ in order to demonstrate the cyclic four-spin exchange dependence of the heat capacity more clear and straightforward, however, the item $\max(0,\mathcal{C})$ shows all behaviors of $\mathcal{C}$.
We observe the heat capacity decreases suddenly when the controlling parameter $K$ is tuned towards the critical point $K=-\mathcal{J}_{Is}$ for various values of the field. Here, we identified this critical point as a singular point for the heat capacity as function of all tunable parameters. At this point the heat capacity has an unconventional behavior. Inset of figure \ref{fig:2DHeatCapacityK} illustrates the heat capacity behavior versus $K/\mathcal{J}_{Is}$ for several values of the coupling constant $\mathcal{J}_{\perp}$ under the considered conditions. It can be seen that the heat capacity singularity with respect to $K/\mathcal{J}_{Is}$ at various values of the magnetic field is directly dependent on the value of the coupling constant $\mathcal{J}_{\perp}$. Indeed, the singularity can be obtained via linear equation $K=-1/2\mathcal{J}_{\perp}$. Therefore, if we want investigate the heat capacity as function of any parameters applied in the ladder Hamiltonian (\ref{SLhamiltonian}) we should consider what is values of the $K$ and $\mathcal{J}_{\perp}$, thus they should be selected so that the singularity does not occur. In result, they should be chosen from the spectrum $K\neq-1/2\mathcal{J}_{\perp}$.

We note that if one wants to investigate any thermal and thermodynamic parameter of an infinite size spin ladder with added cyclic four-spin exchange, he should pay attention to the singularity relation covering $K$ and $\mathcal{J}_{\perp}$ for which, in addition to the heat capacity, all characters such as the magnetization, magnetic susceptibility and the concurrence behave unconventionally near the singular points (in this case near points $K=-1/2\mathcal{J}_{\perp}$). In this regard, we investigated all thermodynamic parameters and thermal concurrence and interestingly concluded that among of them just the heat capacity function in $K-\mathcal{J}_{\perp}$ plane can exactly reveal the singularity relation to us. 
\begin{figure}
\begin{center}
\resizebox{0.45\textwidth}{!}{%
\includegraphics{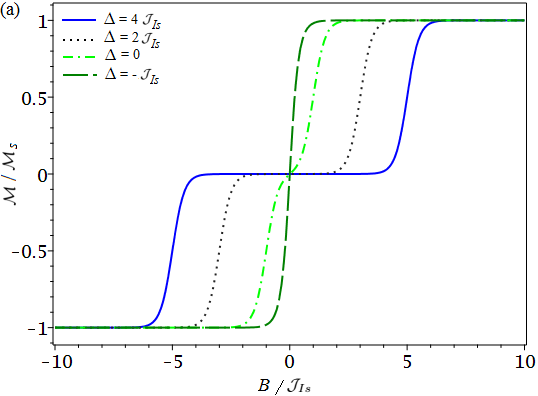}
}
\resizebox{0.45\textwidth}{!}{%
\includegraphics{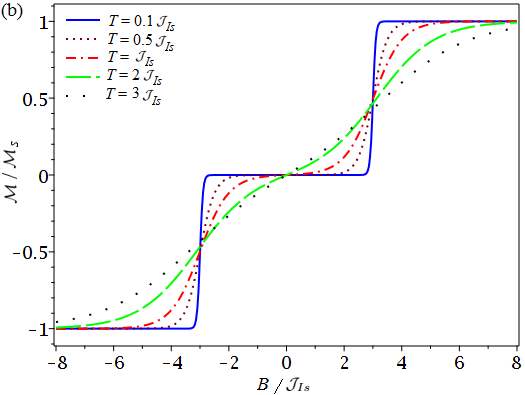}
}
\caption{The magnetic field dependence of the magnetization per block of the spin-1/2 Ising-Heisenberg double sawtooth ladder for various values of (a) the anisotropic parameter $\Delta$ at low temperature $T=0.5\mathcal{J}_{Is}$ and fixed ${J}_{x}=\mathcal{J}_{\parallel}=K=\mathcal{J}_{Is}$, $\mathcal{J}_{\perp}=4\mathcal{J}_{Is}$ (blue solid curve is related to the case $\Delta=4\mathcal{J}_{Is}$), and (b) the temperature $T$ at fixed ${J}_{x}=\mathcal{J}_{\parallel}=K=\mathcal{J}_{Is}$, $\mathcal{J}_{\perp}=4\mathcal{J}_{Is}$, and $\Delta=2\mathcal{J}_{Is}$.}
\label{fig:Magnetization}
\end{center}
\end{figure}

Now let us investigate the magnetization plateaus for the suggested model. The ground-state phase diagram in the presence of an external magnetic field implies various magnetization scenarios depending on the corresponding strength of the Heisenberg and Ising exchange interactions.
For a number of spin models, it was shown that the spin gap existence in the spectrum of magnetic excitations creates plateaus at $\mathcal{M}/\mathcal{M}_{s}\neq 0$ in the external magnetic field.
Also, it was proved that the trimerized spin-1/2 chain exhibits a magnetization plateau at $\mathcal{M}/\mathcal{M}_{s}= 1/3$ which represents a massive phase \cite{Hida}, while the dimerized spin-1 chain \cite{Totsuka} and also mixed-spin (1/2,1) \cite{Sakai} at $\mathcal{M}/\mathcal{M}_{s}= 1/2$ exhibit a plateau (more plateaus were found for spin-1 Heisenberg dimer chain in  \cite{Hovhannisyan1}), where $\mathcal{M}$ is the magnetization and $\mathcal{M}_{s}$ is called saturation magnetization. The magnetization curves with plateaus at $\mathcal{M}/\mathcal{M}_{s}= 0$ and $\mathcal{M}/\mathcal{M}_{s}= 1/3$, also at another values of the magnetization for the Heisenberg spin-1/2 ladders were obtained \cite{Cabra}. The ferromagnetic excitations exhibit a gapless dispersion relation. They reduce the ground-state magnetization, whereas the antiferromagnetic excitations enhance the ground-state magnetization and are gapped from the ground state. Therefore, at low temperature and $B>0$ with regard to the singularity relation between $K$ and $\mathcal{J}_{\perp}$, for instance at the anisotropic point $\Delta=4\mathcal{J}_{Is}$ (solid blue curve in figure \ref{fig:Magnetization}(a)), $\mathcal{M}/\mathcal{M}_{s}$ as a function of ratio $B/\mathcal{J}_{Is}$ jumps down to -1 from 0  when the applied field be $B\approx-5\mathcal{J}_{Is}$,  and it remains unchanged until the field reaches the antiferromagnetic excitation gap $B\approx 5\mathcal{J}_{Is}$ at which it jumps up to 1. Indeed, the jumps from the plateau $\mathcal{M}/\mathcal{M}_{s}= 0$ occur near threshold magnetic field $B=|5\mathcal{J}_{Is}|$. These studies found a plateau at zero magnetization for the model under the consideration, whose width is given by the spin gap in the otherwise smooth magnetization curve.

Another stimulating difference in magnetization behavior is shown in figure \ref{fig:Magnetization}(b). When the temperature increases, the magnetization plateaus gradually disappear and the magnetization curve will become a smooth curve.  At low temperature, we see that with increase of the anisotropic parameter $\Delta$, the middle plateau at $\mathcal{M}/\mathcal{M}_{s}= 0$ gradually disappear and in turn the threshold magnetic field at which the magnetization jumping occurs will decrease (see  figure \ref{fig:Magnetization}(a) again). This means that by tuning the Heisenberg anisotropic parameter in the spin ladder, the threshold magnetic field at which phase transition occurs will change.

\begin{figure}
\begin{center}
\resizebox{0.6\textwidth}{!}{%
\includegraphics{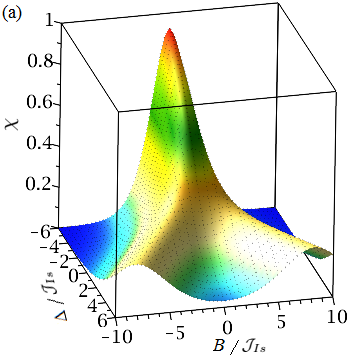}
\includegraphics{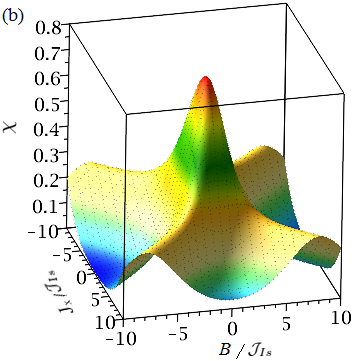} 
}
\resizebox{0.3\textwidth}{!}{
\includegraphics{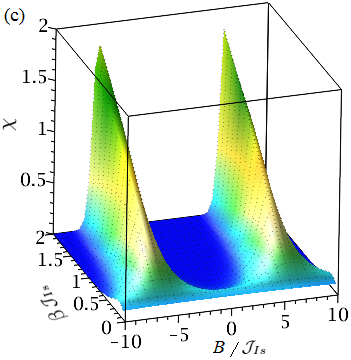} 
}
\caption{Magnetic susceptibility at temperature $\beta=0.25\mathcal{J}_{Is}^{-1}$ as a function of (a) $B/\mathcal{J}_{Is}$ and $\Delta/\mathcal{J}_{Is}$ for fixed $\mathcal{J}_{\parallel}=K={J}_{x}=\mathcal{J}_{Is}$ and $\mathcal{J}_{\perp}=4\mathcal{J}_{Is}$, (b) $B/\mathcal{J}_{Is}$ and ${J}_{x}/\mathcal{J}_{Is}$ for fixed $\mathcal{J}_{\parallel}=K=\mathcal{J}_{Is}$, $\mathcal{J}_{\perp}=4\mathcal{J}_{Is}$ and $\Delta=-2\mathcal{J}_{Is}$. (c) Magnetic susceptibility as function of the temperature parameter $\beta\mathcal{J}_{Is}$ and magnetic field $B/\mathcal{J}_{Is}$ at fixed values of ${J}_{x}=\mathcal{J}_{\parallel}=K=\mathcal{J}_{Is}$, $\mathcal{J}_{\perp}=4\mathcal{J}_{Is}$ and $\Delta=4\mathcal{J}_{Is}$.}
\label{fig:3DMS}
\end{center}
\end{figure}
The effects of anisotropic parameter $\Delta$ and XX coupling ${J}_{x}$ on typical thermal variations of the magnetic susceptibility versus field $B/\mathcal{J}_{Is}$ are displayed in figure \ref{fig:3DMS} under the conditions $\mathcal{J}_{\parallel}=K=\mathcal{J}_{Is}$. Figure \ref{fig:3DMS}(a) shows the magnetic susceptibility with respect to the $\Delta/\mathcal{J}_{Is}$ and field $B/\mathcal{J}_{Is}$ at  $\beta=0.25\mathcal{J}_{Is}^{-1}$, where the coupling constants are taken as $\mathcal{J}_{\parallel}=K={J}_{x}=\mathcal{J}_{Is}$ and $\mathcal{J}_{\perp}=4\mathcal{J}_{Is}$. Here, regions of the states CFM and CFI (blue regions) are separated through a peak of zero-field susceptibility (which ground state is EFI 1 state) at negative values of the anisotropic parameter $\Delta$. With increasing the ratio $\Delta/\mathcal{J}_{Is}$, the peak range decreases until at $\Delta/\mathcal{J}_{Is}>0$ it is divided into two smaller peaks such that one of them appears in the range of $B/\mathcal{J}_{Is}>0$ and the other in the range of $B/\mathcal{J}_{Is}<0$. Clearly, first peak (in the negative filed) depicts that the ground state is the EAFM state (see figure \ref{fig:QPTBDelta}), while the second peak depicts that the ground state is the CFI state. Minimum between two peaks presents the state EFI 2 region.

More features of the magnetic susceptibility is illustrated in figure \ref{fig:3DMS}(b). The symmetry of the susceptibility diagram is explicit versus both axes $B/\mathcal{J}_{Is}$ and ${J}_{x}/\mathcal{J}_{Is}$ for the conditions $\mathcal{J}_{\parallel}=K=\mathcal{J}_{Is}$, $\mathcal{J}_{\perp}=4\mathcal{J}_{Is}$ and $\Delta=-2\mathcal{J}_{Is}$. As heat capacity contour plot presented in figure  \ref{fig:heatcapacitycontour}(b), here the phase regions are obviously identifiable. It can be seen that at zero the coupling constant ${J}_{x}$, we have maximum zero-field magnetic susceptibility (biggest peak at the middle of the diagram). This peak represents that the ground state of the system under the considered conditions is EAFM state, which is general feature of quantum antiferromagnetically coupled spin ladder. When absolute values of the XX dimer coupling constant $\vert{J}_{x}/\mathcal{J}_{Is}\vert$ and the field $\vert B/\mathcal{J}_{Is}\vert$ increase, the peak is divided into four identical smaller peaks, whose with further increase of the both parameters, they are moved away from each other. These small peaks specify phase boundaries illustrated at figure \ref{fig:QPTBJ}. Then, minimums between two peaks at the negative values of the ${J}_{x}/\mathcal{J}_{Is}$ and the positive values of the ${J}_{x}/\mathcal{J}_{Is}$ represent regions of the state EFI 2. Blue regions at the negative and positive values of the field and finite values of $\vert{J}_{x}/\mathcal{J}_{Is}\vert$ represent , respectively, the states CAF-EFI 1 and the state CFI.

Figure \ref{fig:3DMS}(c) depicts the magnetic susceptibility as function of the temperature parameter $\beta\mathcal{J}_{Is}$ and the magnetic field $B/\mathcal{J}_{Is}$ at fixed values of  ${J}_{x}=\mathcal{J}_{\parallel}=K=\mathcal{J}_{Is}$, $\mathcal{J}_{\perp}=4\mathcal{J}_{Is}$ and $\Delta=4\mathcal{J}_{Is}$. By inspecting this figure and figure \ref{fig:Magnetization}(a) (blue solid curve) simultaneously, one can find that near the absolute threshold magnetic field at which the magnetization jumps occur the magnetic susceptibility reaches its maximum. It can be interestingly seen that along the plateau $\mathcal{M}/\mathcal{M}_{s}= 0$ which the magnetization does not change, the magnetic susceptibility is zero and remains unchanged as well (review figure \ref{fig:3DMS}(c) at range $B<|5\mathcal{J}_{Is}|$ when $\beta\mathcal{J}_{Is}=2$). 
\section{Correlation function and thermmal concurrence}\label{Concurrence}
Due to the translation invariance and $U(1)$ invariance $\big[H_{SL},\sum_{k=1}^N\sigma_k^z\big]=0$ of the spin ladder Hamiltonian, the reduced density matrix of the nearest neighbor two qubit in the presence of the magnetic field can be expressed as follows, 
\begin{equation}\label{density matrices}
\mathbf{\rho_i} = \left(
\begin{array}{cccc}
u_+ & 0 & 0 & 0 \\
0 & w_1 & z^* & 0 \\
0 & z & w_2 & 0\\
 0 & 0 & 0 & u_-
\end{array} \right),
\end{equation}
where 
\begin{equation}\label{Correlation}
\begin{array}{lcl}
u_{\pm}= \frac{1}{4}(1+\langle\sigma_{i,2}^z\sigma_{i,5}^z\rangle)\pm \frac{\langle\sigma_{i,2}^z+\sigma_{i,5}^z\rangle}{2},\\
w_1=w_2= \frac{1}{4}(1-\langle\sigma_{i,2}^z\sigma_{i,5}^z\rangle),\\
z=z^*=\frac{1}{2}\langle\sigma_{i,2}^x\sigma_{i,5}^x\rangle.
\end{array}
\end{equation}
The correlation function between Heisenberg spin dimers when $N\rightarrow \infty $ can be obtained by using a derivative of the free energy with respect to the related parameters as 
\begin{equation}\label{Correlationfunctions}
\begin{array}{lcl}
\big\langle\sigma^x_{2}\sigma^x_{5}\big\rangle=-\frac{1}{2}\frac{\partial f}{\partial{J}_x},\\
\big\langle\sigma^z_{2}\sigma^z_{5}\big\rangle=-\frac{\partial f}{\partial\Delta},\\
\mathcal{M}_i=\big\langle\sigma^z_i\big\rangle=-\frac{\partial f}{\partial B}.
\end{array}
\end{equation}
Where on a block $\mathcal{M}= \langle\sigma_{2}^z+\sigma_{5}^z\rangle/2$ is the magnetization.  

Here, we would like apply two correlation functions (\ref{Correlationfunctions}) between the spins from the same Heisenberg dimer in a block and the corresponding magnetization in order to calculate the concurrence. After some algebraic manipulation the concurrence will be as the form \cite{Osterloh1}
\begin{equation}\label{Tm}
\begin{array}{lcl}
C_{2,5}(i)=\\
2\max\big\lbrace 0,\vert\langle\sigma_{i,2}^x\sigma_{i,5}^x\rangle\vert-\frac{1}{2}\vert 1+
\sqrt{\big(\langle\sigma_{i,2}^z\sigma_{i,5}^z\rangle-\mathcal{M}\big)\cdot\big(\langle\sigma_{i,2}^z\sigma_{i,5}^z\rangle+\mathcal{M}\big)+\frac{1}{2}\mathcal{M}+\frac{1}{16}}\vert\big\rbrace
\end{array}
\end{equation}

\begin{figure}
\begin{center}
\resizebox{0.6\textwidth}{!}{%
\includegraphics{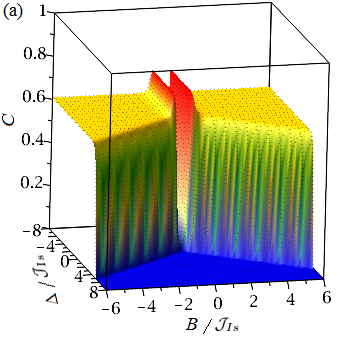}
\includegraphics{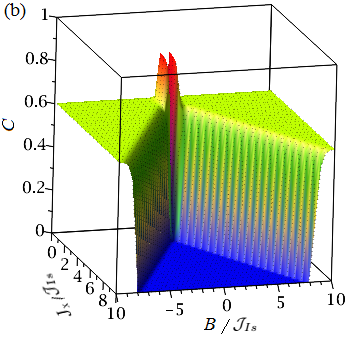}
}
\resizebox{0.4\textwidth}{!}{
\includegraphics{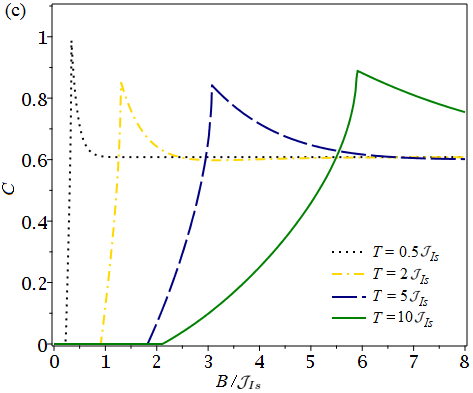} 
}
\caption{Concurrence at low temperature $T=0.5\mathcal{J}_{Is}$ as function of  (a) $B/\mathcal{J}_{Is}$ and $\Delta/\mathcal{J}_{Is}$, for fixed $\mathcal{J}_{\parallel}=K={J}_{x}=\mathcal{J}_{Is}$ and $\mathcal{J}_{\perp}=4\mathcal{J}_{Is}$, and (b) $B/\mathcal{J}_{Is}$ and ${J}_{x}/{J}_{Is}$, for fixed $\mathcal{J}_{\parallel}=K=\mathcal{J}_{Is}$, $\mathcal{J}_{\perp}=4\mathcal{J}_{Is}$ and $\Delta=-2\mathcal{J}_{Is}$. (c) Magnetic field variations of the concurrence for various fixed values of the temperature at fixed ${J}_{x}=\mathcal{J}_{\parallel}=K=\mathcal{J}_{Is}$, $\mathcal{J}_{\perp}=4\mathcal{J}_{Is}$ and $\Delta=-2\mathcal{J}_{Is}$.}
\label{fig:2Dconcurrence1}
\end{center}
\end{figure}

The concurrence represents a feasible measure of bipartite entanglement for quantum spin systems. To elucidate the effect of magnetic field upon the bipartite entanglement, we have plotted in figure \ref{fig:2Dconcurrence1} the concurrence as function of the magnetic field, the anisotropic parameter $\Delta$, isotropic parameter ${J}_{x}$ and temperature $T$. As shown in figure \ref{fig:2Dconcurrence1}(a), in the absence of the magnetic field, the concurrence is minimum along the axis $\Delta/\mathcal{J}_{Is}$. With increasing the absolute value of the field from zero, the concurrence arises and reaches its maximum. With further increase of the magnetic field absolute value, the concurrence gradually decreases and finally reaches plateaus for which the concurrence does not change for $\Delta<0$, but for the case when $\Delta>0$ and $|B|>0$, by increasing the anisotropic parameter $\Delta$ the concurrence decreases and suddenly vanishes. At high ranges of $\Delta$ the concurrence remains disappear.

Figure \ref{fig:2Dconcurrence1}(b) gives the thermal concurrence as a function of the magnetic field and the isotropic parameter ${J}_{x}$ at low temperature and fixed $\Delta=-2\mathcal{J}_{Is}$. In this case, the behavior of the concurrence is almost similar to the previous case, namely, in the presence of an external weak magnetic field for finite values of ${J}_{x}$ the concurrence becomes maximum ($C=1$). With increase of ${J}_{x}$ the concurrence suddenly decreases and disappears at a critical point of ${J}_{x}$. When the absolute value of the magnetic field increases, the range of this critical point increases, hence, the entanglement vanishing occurs at the stronger isotropic coupling constant ${J}_{x}$.

In figure \ref{fig:2Dconcurrence1}(c) we display the concurrence as a function of the magnetic field for various values of the temperature, assuming fixed values of the anisotropic parameter $\Delta=-2\mathcal{J}_{Is}$, $K=\mathcal{J}_{Is}$, coupling constants $\mathcal{J}_{\parallel}={J}_{x}=\mathcal{J}_{Is}$ and $\mathcal{J}_{\perp}=4\mathcal{J}_{Is}$. It is explicit when the magnetic field is decreased the concurrence  decreases and finally vanishes at a threshold magnetic field. With increase of the temperature, the range of the threshold magnetic field gradually increases.

\begin{figure}
\begin{center}
\resizebox{0.4\textwidth}{!}{%
\includegraphics{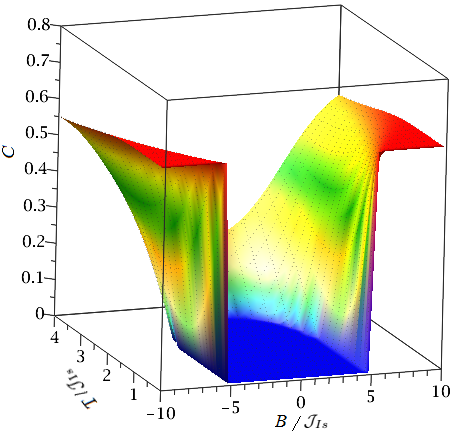}
}
\caption{Concurrence as function of the temperature $T/\mathcal{J}_{Is}$ and $B/\mathcal{J}_{Is}$ for the fixed $\mathcal{J}_{\parallel}=K={J}_{x}=\mathcal{J}_{Is}$, $\mathcal{J}_{\perp}=4\mathcal{J}_{Is}$ and $\Delta=4\mathcal{J}_{Is}$.}
\label{fig:2Dconcurrence2}
\end{center}
\end{figure}

Figure \ref{fig:2Dconcurrence2} shows the concurrence as function of the temperature and the magnetic field for fixed $\mathcal{J}_{\parallel}={J}_{x}=\mathcal{J}_{Is}$ and $\mathcal{J}_{\perp}=4\mathcal{J}_{Is}$, where with regard to figures \ref{fig:Magnetization}(a) and \ref{fig:3DMS}(c) the anisotropic parameter is taken as $\Delta=4\mathcal{J}_{Is}$.  As shown in figure \ref{fig:2Dconcurrence2}, in the range $|B|<5$, the entanglement disappears at low temperature. This phenomenon is in accordance with the magnetic susceptibility disappearing as we mentioned in figure \ref{fig:3DMS}(c), also with the magnetization plateau at $\mathcal{M}/\mathcal{M}_{s}= 0$ (blue solid cure plotted in figure \ref{fig:Magnetization}(a)). On the other hand, the concurrence is maximum at low temperature in the presence of a magnetic field with magnitude $|B|\gtrsim 5$, in this condition the magnetic susceptibility has two peaks. We here remind that as we mentioned before, the magnetization jumps occur at this range of the field.
Due to the thermal fluctuations gradually destroy the quantum entanglement, hence, when the temperature increases, the concurrence generally decreases.

Therefore, we conclude that at low temperature the behaviors of the all three functions $\mathcal{M}/\mathcal{M}_{s}$, $\chi$ and $C$  mach each other versus the magnetic field changes under the considered conditions.
\section{Conclusions}\label{conclusion}
In this work, we have exactly solved a spin-1/2 double sawtooth ladder included interstitial Heisenberg dimer spins connected to the leg spins on each block, where four spins on the square plaquette have cyclic four-spin exchange interaction by taking advantage of the transfer-matrix direct formalism. The spin ladder is obtained from distorting a diamond-like spin chain. The considered model shows the full block separability, hence it is capable of being treated as an exactly solvable quantum model. The elaborated rigorous procedure has enabled us to obtain exact results for the ground-state phase diagram and lead to the possibility of obtaining exact analytical expressions for the thermodynamic quantities such as the heat capacity, Gibbs free energy, the magnetization, the magnetic susceptibility and pair correlation functions. 

First, the heat capacity have been presented for special values of the coupling constants $\mathcal{J}_{\parallel}$, ${J}_{x}$, $\mathcal{J}_{\perp}$ and $\Delta$, also cyclic four-spin exchange interaction $K$. In result, this function showed the boundaries between the ground states of the model with high accuracy. Moreover, we found that between all thermodynamic parameters, the heat capacity is sensitive versus the changes of the cyclic four-spin exchange $K$. To gain an overall insight into the heat capacity behavior in the presence of a magnetic field at low temperature, we explored in detail the cyclic four-spin exchange interaction dependence of the heat capacity and concluded that this function depicts a singularity versus $K/\mathcal{J}_{Is}$, which can be defined as a linear equation of $K$ and $\mathcal{J}_{\perp}$ , i.e., $K=-1/2\mathcal{J}_{\perp}$. Then if one wants investigate the thermodynamic parameters of the system and also thermal pairwise entanglement of the interstitial Heisenberg dimer, he should pay attention to those values of the cyclic four-spin exchange interaction which satisfy the singularity relation $K=-1/2\mathcal{J}_{\perp}$. In order words, for realizing the behavior of thermodynamic parameters and thermal entanglement quantities in high accuracy, we should select the values of the $K$ and $\mathcal{J}_{\perp}$ from spectrum $K\neq -1/2\mathcal{J}_{\perp}$.

The magnetization curve versus the ratio $|B/\mathcal{J}_{Is}|$ has been investigated in detail. At low temperature we observed a plateau at $\mathcal{M}=0$ which by decreasing the Heisenberg anisotropy parameter $\Delta$ this plateau is gradually disappeared. Indeed, we found that for the cases where the zero magnetization plateau is appear, the threshold magnetic field range at which the magnetization jumping occurs, is decreased by decreasing the anisotropic parameter $\Delta$. 

As well as the heat capacity, we here investigated the magnetic susceptibility with respect to the magnetic field for several values of the anisotropic parameter $\Delta$ and also XX Heisenberg dimer coupling constant ${J}_{x}$, where with regard to the singularity relation, other parameters applied in the Hamiltonian of the system under the consideration were taken as fixed values. It is quite surprising that  the minima and maxima of this function also represent the ground states regions with high accuracy specially in the case where the ground state is the EAFM state that is corresponds to the highest peak at the region around the zero magnetic field and zero ${J}_{x}$. 

Furthermore, we presented a remarkable characteristic of quantum entanglement as given by the concurrence. 
Even when cyclic four-spin exchange interaction does not contribute directly in quantum entanglement between interstitial Heisenberg dimer in each block (except when it is a singular point as discussed before), we discussed the thermal entanglement within the dimer spins of the Ising-Heisenberg double sawtooth spin ladder structure and the threshold magnetic field. First we obtained the Heisenberg dimer operators immersed in the ladder. Thereafter, the concurrence is obtained straightforwardly in terms of the reduced density matrix operator elements. Through the correlation functions, we obtained the elements of the reduced density matrix by using the free energy per block when the size of the system is taken as $N\rightarrow \infty$ in high accuracy. Using the concurrence, we studied the pairwise entanglement for the interstitial Heisenberg dimer of the model with added cyclic four-spin exchange interaction between spins on the square plaquette of each block as a function of Hamiltonian parameters. At low temperature and zero magnetic field the entanglement disappears. In the presence of an external magnetic field ($|B|>0$) the concurrence is maximum at low temperature in the considered conditions. By increasing exchange coupling parameters of the interstitial Heisenberg dimer, the concurrence decreases and reaches its minimum ($C=0$), here, the state of the interstitial Heisenberg dimer is not entangled. 
Moreover, we observed that at low temperature the concurrence has a threshold magnetic field at which the phase transition occurs for the system. Generally, the threshold magnetic field range changes with change of the $\Delta$, ${J}_{x}$ and $T$ under the considered conditions. 

To demonstrate the accuracy of our statements, as an example, we investigated the magnetic variations of three functions, the magnetization, the magnetic susceptibility and the thermal concurrence between spins of the interstitial Heisenberg dimer at low temperature $T=0.5\mathcal{J}_{Is}$, where the applied parameters in the Hamiltonian were taken as ${J}_{x}=\mathcal{J}_{\parallel}=K=\mathcal{J}_{Is}$, $\mathcal{J}_{\perp}=4\mathcal{J}_{Is}$ and $\Delta=4\mathcal{J}_{Is}$. We concluded that the behaviors of all three functions mach each other, namely, for $B\approx \pm 5\mathcal{J}_{Is}$  the concurrence and the magnetic susceptibility become maximum, which it is in accordance with the magnetization jumping near the threshold magnetic field. In the range $5\mathcal{J}_{Is}<B<5\mathcal{J}_{Is}$ at low temperature, when the magnetization has a plateau at  $\mathcal{M}=0$, both functions, the magnetic susceptibility and the concurrence disappear.

We consider tunable parameters in the Hamiltonian of the suggested double sawtooth ladder because a tunable sawtooth spin ladder would offer a versatile platform for the study on the spin interaction of quantum many-body states with ultracold atoms. As one of the strong applications of the double sawtooth ladder considered in this paper, we can point out to the quantum state transmission through quantum spin systems such as our model which surely can be useful in quantum information processing, spintronics, optical lattices and theoretical and experimental condensed matter physics.

Finally, it should be also mentioned that the precise method elaborated in the present paper can be straightforwardly adapted to account for the mixed spin (1/2,1) Ising-Heisenberg ladder as well. Our future work will continue in this direction.
\section{Acknowledgments}
NA acknowledge partly financial support by the RA MES State Committee of Science, in the frames of the research project No. SCS 15T-1C114 and the MC-IRSES no. 612707 (DIONICOS) under FP7-PEOPLE-2013.

 {
\end{document}